\documentclass[10pt,aps,prd,superscriptaddress,twocolumn,amsmath,amssymb,floatfix, longbibliography,
nofootinbib 
]{revtex4-1}

\usepackage[dvips]{graphics}
\usepackage{bm}
\usepackage{epsfig}
\usepackage{enumerate}
\usepackage{subfigure}
\usepackage{color}
\usepackage{graphicx}
\usepackage[dvipsnames]{xcolor}
\usepackage{graphicx}
\usepackage[colorlinks,citecolor=blue,linkcolor=blue,urlcolor=blue,plainpages=false,pdfpagelabels]{hyperref}

\newcommand{\beq}{\begin{equation}}
\newcommand{\eeq}{\end{equation}}
\newcommand{\bea}{\begin{eqnarray}}
\newcommand{\eea}{\end{eqnarray}}

\begin{document} 
\title{Theory of light diffusion through amplifying photonic lattice} 

\author{SK Firoz Islam}
\affiliation{Department of Applied Physics, Aalto University, P.~O.~Box 15100, FI-00076 AALTO, Finland}

\author{Alexander A. Zyuzin}
\affiliation{Department of Applied Physics, Aalto University, P.~O.~Box 15100, FI-00076 AALTO, Finland}
\affiliation{Ioffe Physical--Technical Institute,~194021 St.~Petersburg, Russia}

\begin{abstract}
We present a study of radiation propagation through disordered amplifying honeycomb photonic lattice,
where elastic scattering provides feedback for light generation.
To explore the interplay of different scattering mechanisms and the amplification background, 
we consider the Dirac Hamiltonian with a random potential and derive diffusion equation for the average intensity of light.
The transmission coefficient and interference correction to the diffusion coefficient are enhanced near the lasing threshold. 
The transition between weak anti-localization and weak localization behaviours might be controlled by the parameters associated
with the amplification and inter-valley scattering rates. 
\end{abstract}
\maketitle
\section{Introduction}
Disordered amplifying optical media have received much attention, most particularly for their diverse applications as random 
lasers \cite{letokhov1968generation,turitsyn2010random, wiersma2013,wiersma2008physics,cao2003lasing}, spatial light confinement
or coherent control \cite{PhysRevLett.84.5584,riboli,PhysRevLett.112.133903,PhysRevApplied.12.064045}, and application to medical technology \cite{yun2017light,polson2004random}. 
The wave interference processes leading to weak localization effects \cite{doi:10.1063} has been studied extensively  with and without
amplification backgrounds
\cite{wiersma1997,PhysRevB.55.5736,Stephen,PhysRevE.51.5274,PhysRevB.56.12038,Zyuzin_1994,PhysRevB.89.224202,PhysRevE.54.4256,PhysRevB.52.7960,PhysRevB.50.9644} in
three dimensional optical media, followed by a number of experiments \cite{deOliveira:97,deOliveira:96,PhysRevA.64.063808,PhysRevLett.93.263901,PhysRevLett.75.1739}.

Recently, noting the analogy to the topological phases in electronic systems \cite{volovik_book, RevModPhys.82.3045}, photonic lattices
(PhL) with topologically nontrivial band structures have generated a lot of interest  \cite{PhysRevLett.100.013904}. In particular, the 
light passing through two dimensional (2D) PhL formed by a triangular periodic array of parallel waveguides can be described by a Dirac
equation and satisfy conical dispersion relation \cite{PhysRevB.44.8565}.
The transmission probability of light through disordered honeycomb PhL has been studied both theoretically and experimentally \cite{PhysRevA.75.063813, Wang_2014}.
In particular, it was shown that the transmission of light with frequencies near the Dirac point is inversely proportional to the length of the sample. 
Moreover, the suppression of the coherent backscattering of radiation at an angle that is reciprocal to the angle of incidence has been theoretically
predicted in Ref. \cite{Sepkhanov_2009}. The weak antilocalization is due the destructive interference of waves caused by the accumulation of $\pi$
Berry phase along a closed trajectory. This effect is similar to the absence of the backscattering of Dirac fermions in graphene \cite{Ando_Nakanishi, Suzuura_Ando}.
Various Lifshitz phase transitions might be achieved in topological PhL by means of tuning the magnetic permeability and dielectric 
permittivity tensor components \cite{PhysRevLett.100.013904}. Photonic lattices with gain or loss are also known for being an ideal platform for investigating the physics of non-Hermitian Hamiltonians \cite{Bender_sense, Berry}. 
For recent reviews on the topological photonics, see Refs.~[\onlinecite{Review_photonics_2018,Alexander,Konotop_RevModPhys}]. 
Although despite intense research on light propagation though photonic lattices, to the best of our knowledge, the combined effects of disorder scattering and
amplification of light in honeycomb PhL has not been considered. 

In this work, we investigate light transmission through disordered amplifying honeycomb photonic lattice. We first derive the diffusion 
equation for the field-field correlation function in the situation where only one of two inequivalent valleys in the photonic band-structure 
is excited by the incident radiation. It is shown that the average transmission is enhanced at the vicinity of the lasing threshold. We also
investigate the interference correction to the transmission of light taking into account the inter-valley scattering processes. It is shown
that that the sign of the correction depends on the parameters of amplification and inter-valley scattering rates. 

\section{Model of amplifying media}\label{sec2}
\begin{figure}[t]
\begin{tabular}{cc}
\includegraphics[width=6cm]{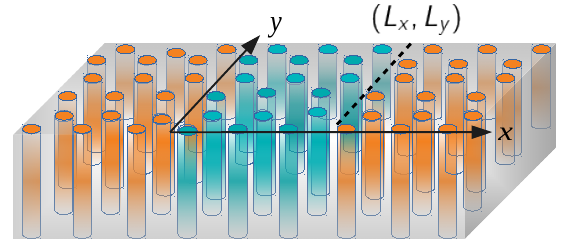}
\includegraphics[width=2cm]{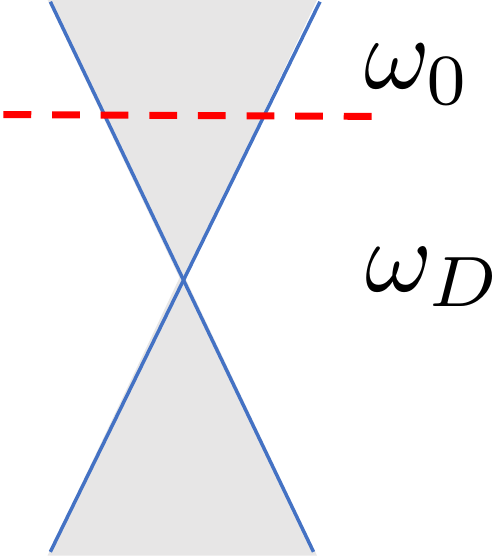}
\end{tabular}
\caption{Left: schematic sketch of the photonic lattice, consists of a disordered amplifying region of length $L_x$ and width $L_y$ placed between two dissipative regions. 
Photonic lattice slab is assumed to have fully transparent boundaries for the radiation at $x=\{0, L_x\}$ and fully reflecting boundaries at $y=\{0, L_y\}$. Right: 
the frequency $\omega$ of the incident light is at the vicinity of the resonance frequency $\omega_0$ above the frequency of the Dirac point $\omega_D$.}
\label{PhGraphene}
\end{figure}

We consider a 2D PhL in the $x-y$ plane with length $L_x$ and width $L_y$, made of paralleled waveguides aligned along the $z$-axis, as
schematically shown in Fig. \ref{PhGraphene}. The waveguides are arranged in a manner to form a graphene-like hexagonal lattice, replacing each sublattice atoms of graphene. 
The waveguides and the environment media between them are described by a frequency dependent dielectric permittivity $\epsilon_{\omega}(x,y)$. 
To specify, the dielectric permittivity of environment is assumed to be real and positive in all scattering regions. 
The permittivity of waveguides is real and positive in the regions $x<0$ and $x>L_x$ as well. Although, in the middle region $0<x<L_x$ the permittivity
is complex, for which we adopt the model of oscillating electric dipole response at the resonance frequency $\omega_0$ and in the situation with inversion
population.

We consider in-plane propagation of the $E_z(x,y,t)\sim E_z(x,y,\omega)e^{-i\omega t}$ component of the TE-polarized electric field with frequency $\omega$.
As it was shown, for example in Ref. \cite{PhysRevB.98.165129}, the field component $E_z(x,y, \omega)$ on the honeycomb lattice can be written via 
irreducible doublet and singlet representations for two sets of inequivalent corners of the hexagonal first Brillouin
zone $E_z(x,y, \omega) \rightarrow [E_1(\mathbf{K}/\mathbf{K}', x,y, \omega), E_2(\mathbf{K}/\mathbf{K}', x,y, \omega)],~ E_3(\mathbf{K}/\mathbf{K}', x,y, \omega)$. 
The doublet states form two inequivalent Dirac points at frequency $\omega_D$, while non-degenerate singlet states are separated in frequency from the Dirac points 
and might be eliminated. 

Note that in the amplifying system, in addition to the structure-dependent frequency $\omega_D$, the gain is characterized by the resonance frequency 
of two-level systems $\omega_0$.
We will consider the propagation of EM wave with frequency $\omega \approx \omega_0$ and assume that $\omega_0 > \omega_D$, as schematically shown in 
the right panel of Fig. \ref{PhGraphene}.
To this end, the wave equation reduces to a Dirac equation in subspace of the doublet states for two valleys in the presence of the non-Hermitian background.

Experimentally realistic systems might contain random scattering processes due to waveguide lattice imperfections that can play a key role in the light transport through the PhL.
We consider disorder in the region $0<x<L_x$ by adding a small fluctuation to the real part of the dielectric constant, while keeping imaginary part constant. 
Generally, disorder allows for both inter and intra-valley (corresponding to two inequivalent Dirac points of the honeycomb lattice) scattering processes.
Although, we will first neglect inter-valley scattering and consider a situation in which incident EM field excites only a single Dirac valley. 
We shall comment on the other pumping mechanisms and scattering channels later in the section devoted to the light interference effects.

In this model, equation for the field $(E_1, E_2)$ (with the wave-vector expanded in the vicinity of one of the two inequivalent Dirac points), 
with a random scalar potential $V(\mathbf{r})$ is given by
\begin{equation}\label{Disorder_1}
\frac{\omega_D}{c^2}
\bigg\{
\left[
\tilde{\omega} + i\Gamma - V(\mathbf{r})
\right]\sigma^0_{ab} +iv\boldsymbol{\sigma}_{ab}\cdot \boldsymbol{\nabla} 
 \bigg\}
 E_b(\mathbf{r},\omega)=0, 
\end{equation}
where $\mathbf{r} = (x, y)$ is the position vector, $\sigma^0\equiv 1_{2\times2}$,
$\boldsymbol{\sigma} = (\sigma^{x}, \sigma^{y})$ is the vector composed of Pauli matrices acting in the subspace of the doublet states at a corner of the crystal Brillouin zone, and $v$ is the group velocity. 
The frequency
\begin{equation}
\tilde{\omega} = \frac{\omega^2-\omega_D^2}{\omega_D}>0
\end{equation}
is introduced for brevity. The term $\Gamma$ is related to the imaginary part of the dielectric permittivity of the cylinders and assumed to be frequency independent. The system with gain is described by the negative sign of $\Gamma$, which will be considered in what follows. Note that the term $\omega_D/c^2$ in Eq.~(\ref{Disorder_1}) is kept to preserve the correct dimensionality of the wave-equation. It is also worthwhile to mention that TM mode satisfies similar equation to Eq. (\ref{Disorder_1}) except different values of the frequency of the Dirac point $\omega_D$ and velocity term $v$. In present work, we proceed with TE mode only.

Random potential with a unit matrix in both the valley and the doublet spaces is assumed to satisfy the following properties
$
 \langle V({\bf r})\rangle=0
$ 
together with 
\begin{equation}
(\omega_D^2/c^4)\langle V({\bf r}) V({\bf r'})\rangle=\Lambda\delta({\bf r-r'}), 
\end{equation}
where the angular brackets denote averaging over the realizations of disorder, and 
$\Lambda = (\omega_D/c^2)/\pi\nu\tau_0$
is related to the mean free time of the radiation due to scattering on impurities $\tau_{0}$ and the density of states per valley and per doublet $\nu = \tilde{\omega}c^2/2\pi v^2\omega_D$.

Finally, we assume that condition
$ 
\tilde{\omega}\tau_0 \gg 1
$ holds
and proceed with the standard diagrammatic technique for the disordered system \cite{Alt_Aronov,montaumbaux}. 

To note the interplay between the scattering and amplification processes, it is instructive to comment on the disorder averaged retarded Green function of Eq. \ref{Disorder_1}, which is given by
\begin{eqnarray}\label{henkel}\nonumber
 \mathcal{D}_{ab}(\mathbf{r},\omega) &=& \sqrt{\frac{\nu c^2}{ 4 v r \omega_D}}\left(\sigma^0_{ab}+\boldsymbol{\sigma}_{ab}\cdot \mathbf{r}/r\right) \\
 &\times& \exp\left[\frac{ir}{v}\left(\tilde{\omega}+\frac{i}{2\tau}\right) + \frac{i\pi}{4}\right].
\end{eqnarray}
Compared with the Green function of electrons in graphene, here the relaxation rate is defined by both random scattering on disorder and amplification as
\begin{equation}
\frac{1}{\tau}=\frac{1}{\tau_0}-\frac{1}{\tau_{A}},
\end{equation}
where $\tau_{A} = 1/|\Gamma|$ is the amplification time. Note that approximation $\tau_0/\tau_{A} < 1$ is considered so that light experiences multiple scattering events on the amplification length \cite{letokhov1968generation}.

\section{Diffusion equation for field-field correlation function}\label{sec3}
Let us briefly review the standard diagrammatic approach to describe the light scattering on random disorder. 
The incoming wave $E_{inc, a}(\mathbf{r},\omega)$ experiences multiple scattering events after entering the media at point $\mathbf{r}$ at the interface $x=0$. 
The propagation of light might be described by the two-point field-field correlation function averaged over disorder
\begin{equation}
\mathcal{I}_{ab}( \mathbf{r},\mathbf{r}';\omega,\omega') = \langle E_{a}( \mathbf{r}, \omega)E^{\dagger}_{b}(\mathbf{r}', \omega')\rangle.
\end{equation}
The correlation function can be obtained by summing up the diffusion ladder
\begin{align}\label{corr2}
 &\mathcal{I}_{bc}( \mathbf{r}_1,\mathbf{r}_2;\omega,\omega') =
 \\\nonumber
 & \Lambda
 \int d\mathbf{r}' d\mathbf{r}''
 \mathcal{D}_{aa'}(\mathbf{r}_1- \mathbf{r}', \omega)\mathcal{D}^{+}_{d'd}(\mathbf{r}_2-\mathbf{r}', \omega') 
 \\\nonumber
 &\times
\mathcal{F}_{a'b;c d'}(\mathbf{r}',\mathbf{r}'',\omega,\omega')E_{inc, a}(\mathbf{r}'', \omega)E^{+}_{inc, d}(\mathbf{r}'', \omega'),
\end{align}
where the integrand term $E_{inc, a}(\mathbf{r}'', \omega)E^{+}_{inc, a}(\mathbf{r}'', \omega')$ describes the intensity of the incident non scattered radiation provided $\omega = \omega'$.
The average intensity of light transmitted though the media is given by $\mathcal{I}_{ab}( \mathbf{r}_1,\mathbf{r}_1;\omega,\omega')$ at position $\mathbf{r}_1 = (L_x, y)$. 

\begin{figure}[t]
\centering
\includegraphics[width=\linewidth]{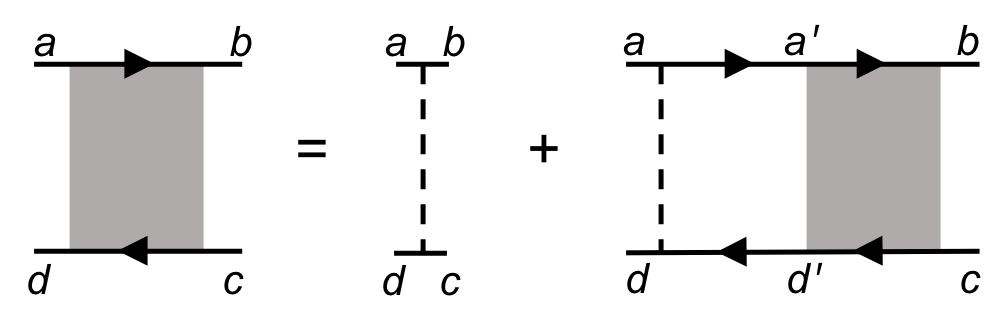}
\caption{Schematic representation of the diffusion ladder in Eq. \ref{Diffusion_1}.}
\label{diffusion}
\end{figure}

The kernel $\mathcal{F}_{ab;cd}(\mathbf{r},\mathbf{r}',\omega,\omega')$ of Eq. \ref{corr2} satisfies diffusion equation, which can be conveniently written in the momentum representation in the form
\begin{align}\label{Diffusion_1}
 &\mathcal{F}_{ab;cd}({\bf q},\omega,\omega')=\delta_{ab}\delta_{cd}
 \\
 &+\Lambda\int\frac{d^2p}{(2\pi)^2}
 \mathcal{D}_{aa'}({\bf p},\omega)\mathcal{D}^{+}_{d'd}({\bf p-q},\omega')
 \mathcal{F}_{a'b;cd'}({\bf q},\omega,\omega'),\nonumber
\end{align}
where $\mathbf{p}=(p_x,p_y)$ and
\begin{equation}
 \mathcal{D}_{ab}({\bf p},\omega)=\frac{c^2}{2\omega_D}\frac{\sigma^{0}_{ab}+\boldsymbol{\sigma}_{ab}\cdot \mathbf{p}/p}{\tilde{\omega}-vp+\frac{i}{2\tau}}
\end{equation}
is the Fourier representation of the Green function Eq. \ref{henkel}.
The diagrammatic representation of the Eq. \ref{Diffusion_1} is shown in Fig.~(\ref{diffusion}). 

To proceed, it is convenient to write the diffusion equation in the new basis as
\begin{equation}\label{new_basis}
F_{\eta \eta'} ({\bf q},\omega,\omega')= \frac{1}{2}\sigma^{\eta}_{da}\sigma^{\eta'}_{bc}\mathcal{F}_{ab;cd}({\bf q},\omega,\omega'),
\end{equation}
where indices $\eta, \eta'$ take the values $\{0,x,y,z\}$.
In this basis the diffusion equation in momentum representation is given by
\begin{eqnarray}
 \left(
 \begin{matrix}\label{diff}
  \frac{Dq^2}{2}-i\delta\omega - \tau^{-1}_{A} &i\frac{v}{2}q_x&i\frac{v}{2}q_y&0
  \\ i \frac{v}{2}q_x&\frac{1}{2}\tau_0^{-1}&0&0\\ i\frac{v}{2}q_y&0&\frac{1}{2}\tau_0^{-1}&0\\0&0&0&\tau_0^{-1}
 \end{matrix}
 \right) F =\frac{1}{\tau}.~~~
\end{eqnarray}
Here $D=v^2\tau$ is the diffusion coefficient and $\delta \omega= 2\omega(\omega-\omega')/\omega_D$. 
The field-field correlation function can be split into singlet $\eta=0$ and triplet $\eta=x,y,z$ components.
The triplet components are suppressed on the scale of the mean free path $\ell =v \tau_0$ at the vicinity of the boundaries and might be eliminated.

As a result, one obtains equation for the remaining singlet component $F_{0}\equiv F_{00}$ in the form
\begin{equation}\label{diff2}
 \left[i\delta\omega+D\boldsymbol{\nabla}_r^2+\frac{1}{\tau_{A}}\right]F_{0}(\mathbf{r},\mathbf{r}';\omega,\omega') =- \frac{1}{\tau}\delta(\mathbf{r}-\mathbf{r}').
\end{equation}
Note the appearance of the positive term $\tau_A^{-1}$ in the diffusion equation for the singlet mode compared with the case of electron diffusion in graphene, \cite{PhysRevLett.97.146805}. Such term signals for the 
size-dependent pole in the diffusion propagator as can be seen after imposing the boundary conditions.

Photonic lattice slab is assumed to have fully transparent boundaries for the radiation at $x=\{0, L_x\}$ and fully reflecting boundaries at $y=\{0, L_y\}$, which is described by
$F_{0}(\mathbf{r},\mathbf{r}';\omega,\omega')|_{x=\{0, L_x\}}  = 0$ and $\partial_y F_0(\mathbf{r},\mathbf{r}';\omega,\omega')|_{y=\{0, L_y\}} = 0$, respectively. 
Hence, the solution of diffusion equation Eq.~(\ref{diff2}) is given by
\begin{align}\nonumber\label{Result_F0}
 &F_{0}(\mathbf{r},\mathbf{r}';\omega,\omega') = \frac{\tau_{A}}{\tau}\sum_{n=1}^{\infty}\sum_{m=0}^{\infty}\frac{u_{n,m}(x,y)u_{n,m}(x',y')}{\lambda_{n,m} -i\delta\omega \tau_{A}},\\\nonumber
 &u_{n,m}(x,y)=\frac{2}{\sqrt{L_xL_y}}\sin\left(\frac{n\pi x}{L_x}\right)\cos\left(\frac{m\pi y}{L_y}\right),\\
  &\lambda_{n,m}= L_{A}^2\left[\frac{n^2}{L_x^2}+\frac{m^2}{L_y^2} \right] - 1,
\end{align}
where $n,m$ are integers and $L_{A}=\pi\sqrt{D\tau_{A}}$ is the critical length, which determines the random amplifier to generator transition. 

The contribution of the lowest mode with $n=1, m=0$ diverges at $L_x=L_A$, where
the Thouless energy $\pi^2 D/L_x^2$ becomes equal to the amplification rate $\tau_{A}^{-1}$. 

\section{Transmission probability}
Let us now proceed to the evaluation of average transmission of radiation incident at an angle $\theta$ on the interface $x=0$ and transmitted at an angle $\phi$ from $x=L_x$. We consider the slab with dimensions $L_x, L_y$ to be smaller than the critical length $L_{A}=\pi\sqrt{D\tau_{A}}$, above which the system becomes a random generator.  We also consider the width of the incident beam to
be larger than $L_y$. 

The transmission coefficient can be defined as a ratio of the transmitted to total incoming flux densities.
The average intensity of radiation with frequency $\omega$ at point $\mathbf{r}$ is given by
\begin{equation}
 I_{\alpha}(\mathbf{r},\omega) =\frac{1}{2}\sigma^{\alpha}_{ab}\mathcal{I}_{ab}( \mathbf{r},\mathbf{r};\omega,\omega).
\end{equation}
Utilizing Eq. \ref{corr2} and Eq. \ref{diff2}, one arrives at the continuity equation in the form
\begin{eqnarray}
\frac{\partial I_0(\mathbf{r},t)}{\partial t} + \boldsymbol{\nabla} \cdot\mathbf{J}_0(\mathbf{r},t) = \frac{I_0(\mathbf{r},t)}{\tau_{A}},
\end{eqnarray}
where the singlet flux density is given by $\mathbf{J}_0(\mathbf{r},t) = - D \boldsymbol{\nabla} I_0(\mathbf{r},t)$.
Consider the incident wave (note that only single valley is being pumped) in the form
\begin{equation}
E_{inc}(\mathbf{r}, \omega) = (1, e^{i\theta})^T \sqrt{I_{inc}/2} e^{i (\mathbf{k}\cdot\mathbf{r} - \omega t)} 
\end{equation}
with intensity $I_{inc}$, frequency $\omega$, and
wave-vector $\mathbf{k} = |\tilde{\omega}|(\cos\theta, \sin\theta)/v$. The incoming flux density is defined as 
$\mathbf{J}_{inc}(\omega) = v(\omega_D/\omega) E^{+}_{inc}( \omega, \mathbf{r})\boldsymbol{\sigma} E_{inc}( \omega, \mathbf{r})$, thus the total incoming flux density reads
\begin{equation}
\langle \hat{x}\cdot\mathbf{J}_{inc}(\omega)\rangle = (\omega_D/\omega)vI_{inc} \int_{-\pi/2}^{\pi/2} \frac{d\theta}{2\pi} \cos\theta.
\end{equation}

By employing the relation between the singlet component of the flux and the wave intensity through a disordered medium
the transmission probability is given by
\begin{eqnarray}
  T_{0} =-\frac{ \pi  \omega D  \cos\phi}{v  \omega_D I_{inc}}\int_{0}^{L_y} \frac{d{y}}{L_y}\frac{\partial}{\partial x} I_{0}({\bf r},\omega)\bigg|_{x=L_x}.~~~~~
\end{eqnarray}
This expression can be further simplified after incorporating Eq.~(\ref{corr2}) and Eq.~(\ref{Diffusion_1}) for the intensity as \cite{PhysRevB.55.5736}
\begin{eqnarray}
  T_{0} = - \frac{ \pi \omega \ell^3 }{ 2 \omega_D L_y}\int_{0}^{L_y} d{y}dy' \frac{\partial^2F_{0}({\bf r},{\bf r}',\omega)}{\partial x\partial x'} \bigg|_{{x'=0 \atop x=L_x} }.~~~~~
\end{eqnarray}
Taking into account Eq. \ref{Result_F0}, estimating $\omega_0/\omega_D \approx 1$, and noting that only $m=0$ contributes, one obtains
\begin{eqnarray}
  T_{0} = \pi^2\frac{ \ell \cos\phi}{L_{A}}\mathrm{cosec}\frac{\pi L_x}{L_{A}},
\end{eqnarray}
where the length $L_x$ is cut by the mean free path $\ell$ from below.
At $L_x \gg L_{A}$ the average transmission is given by the usual expression $\langle T_{0} \rangle =  \ell/L_{x}$,  \cite{Stephen}.
To compare, the transmission probability through the wide $L_y> L_x$ ballistic honeycomb photonic lattice at the Dirac point is proportional
to the ratio $L_y/L_x$, \cite{PhysRevA.75.063813}.
Denoting 
\begin{equation}
L_x=L_{A}(1-\Delta), 
\end{equation}
where $0<\Delta \ll 1$, the average transmission probability through disordered amplifying media at the vicinity of the threshold is given by
$\langle T_{0} \rangle = \ell/ \Delta L_{A}$ \cite{PhysRevB.55.5736}. It is enhanced as $1/\Delta$ compared to the system without amplification. 
Expression for the light transmission is similar to the one in disordered optical media Ref.~\cite{PhysRevB.55.5736}. However, distinctions
are expected in the interference phenomena of light due to the presence of several Dirac cones in the band structure of the honeycomb photonic lattice.

\section{Interference correction }\label{sec4}
The interference phenomenon in disordered electronic and photonic systems has been investigated extensively \cite{montaumbaux}. 
Special interest is in the systems with (pseudo)spin-momentum locking and multiple valley degrees of freedom \cite{Ando_Nakanishi, PhysRevLett.97.146805}.
However, it is known that symmetry-breaking scattering processes might play significant role in the interference effects in the honeycomb lattice.
Particularly in graphene,
it has been shown that the interference correction to the conductivity can tern from positive to negative (i.e. from weak antilocalization to weak localization)
by increasing the strength of the inter-valley scattering processes, \cite{PhysRevLett.97.146805, Suzuura_Ando, PhysRevLett.97.016801}. 
The suppression of the coherent backscattering of radiation from a disordered triangular photonic lattices, where only single valley pumping processes were considered, 
has been addressed in Refs.~\cite{Sepkhanov_2009,Wang_2013}. The accessibility of the valley dependent pumping of light in the honeycomb photonic lattice, where the incident beam splits into two beams (corresponding to two inequivalent valleys), 
has been demonstrated in Ref. \cite{valley_dependent_beams}. Although, the interference correction in amplifying optical media has been addressed a long time ago \cite{Zyuzin_1994}, 
to the best of our knowledge, the analogous effects in amplifying photonic lattices have not been studied.

So far we have focused on the case in which only one of the two inequivalent valleys is excited and considered only scalar disorder described by a unit matrix 
in valley and doublet subspaces. To describe intra and inter-valley scattering mechanisms, following convention as used in the study of weak localization in graphene Ref. \cite{PhysRevLett.97.146805}, 
it is convenient to introduce two sets of $4\times4$ Hermitian matrices: $\vec{\Gamma}=(\Gamma_x,\Gamma_y,\Gamma_z)$ and $\vec{\Xi}=(\Xi_x,\Xi_y,\Xi_z)$ 
in ``pseudospin'' and ``isospin'' space representations, respectively.
Each components are defined as  $\Gamma_x=\Pi_x\otimes\sigma_z,~~\Gamma_y=\Pi_y\otimes\sigma_z,~~\Gamma_z=\Pi_z\otimes\sigma_0,$
and  $\Xi_x=\Pi_z\otimes\sigma_x,~~\Xi_z=\Pi_y\otimes\sigma_y,~~\Xi_z=\Pi_0\otimes\sigma_z$, where $\Pi_{0,x,y,z}$ are Pauli matrices, which act in the valley space.
One can rewrite Dirac equation Eq.~(\ref{Disorder_1}) in the new basis representation utilizing matrices $\Xi$ and $\Pi$, and introduce the intra-valley and inter-valley 
scattering rates, as $\Gamma_{\mathrm{i}}$ and $\Gamma_{z}$, respectively, \cite{PhysRevLett.97.146805}.

It was shown in Ref. \cite{PhysRevLett.97.146805} that the interference correction to the diffusion coefficient $\delta D$ is 
determined by the isospin-singlet pseudospin-singlet and three isospin-singlet pseudospin-triplet components of the field-field correlation function, the Cooperon
\begin{eqnarray}\label{weak}
\frac{\delta D}{D} = \frac{\tau}{2\pi \nu} \int \frac{d^2 \mathbf{r}}{L_xL_y}\left\{ \sum_{j=1}^{3}C^{(j)}_0(\mathbf{r},\mathbf{r})-C^{(0)}_0(\mathbf{r},\mathbf{r})  \right\}.~~~
\end{eqnarray}
Here the pseudospin singlet component $C^{(0)}_0(\mathbf{r},\mathbf{r}')$ is given by the solution (\ref{Result_F0}) of Eq. \ref{diff2}. 
It can be shown that, the pseudospin triplet components $C^{(j)}_0(\mathbf{r},\mathbf{r}')$, with $j=1, 2, 3$, obey Eq. \ref{diff2}, in which one has
to perform a formal substitution of the respective relaxation rates as $\tau_{A}^{-1} \rightarrow \tau_{A}^{-1} - \tau_j^{-1}$, where $\tau_{1,2}^{-1}
= \Gamma_z + \Gamma_{\mathrm{i}}$ and $\tau_{3}^{-1} = 2\Gamma_{\mathrm{i}}$ account for the relaxation rates of the Cooperon triplet components, \cite{PhysRevLett.97.146805}.

Note that the boundary conditions for the Cooperon triplet components depend on the quality of the edge of the photonic lattice.
The sharper boundary provides larger inter-valley scattering rates, resulting in the suppression of the Cooperon triplet components. This  
can be described by the following condition $C^{(1,2,3)}_0(\mathbf{r},\mathbf{r}')=0$ at the corresponding boundary, \cite{PhysRevB.77.193403},
which gives
\begin{align}\nonumber\label{Sharp_Cooperon}
 &C^{(j)}_{0}(\mathbf{r},\mathbf{r}';\omega,\omega') = \frac{\tau_{A}}{\tau}\sum_{n=1}^{\infty}\sum_{m=1}^{\infty}\frac{u_{n,m}(x,y)u_{n,m}(x',y')}{\lambda^{(j)}_{n,m} -i\delta\omega\tau_{A}},\\ \nonumber
 &u_{n,m}(x,y)=\frac{2}{\sqrt{L_xL_y}}\sin\left(\frac{n\pi x}{L_x}\right)\sin\left(\frac{m\pi y}{L_y}\right),\\
  &\lambda^{(j)}_{n,m}= L^2_{A}\left[\frac{n^2}{L_x^2}+ \frac{m^2}{L_y^2}\right]+\frac{\tau_{A}}{\tau_j}- 1.
\end{align}
When the inter-valley scattering times at the boundary are much longer compared to the amplification time, one can use the condition on the free
boundary $\partial_y C^{(1,2,3)}_0(\mathbf{r},\mathbf{r}')|_{y=\{0, L_y\}} = 0$. In this case the solution for the pseudospin
triplet components coincide with Eq. \ref{Result_F0}, again with a formal substitution of the respective relaxation rates.

We are now in the position to calculate the interference correction. 
In the case of sharp boundaries with strong inter-valley scattering processes, using Eq. \ref{Sharp_Cooperon} for triplet and Eq. \ref{Result_F0} 
for the singlet, the integration over coordinates in Eq. \ref{weak} yields the following expression
\begin{eqnarray}\nonumber\label{correction_WL}
\frac{\delta D}{D} &=& \frac{\tau_{A}}{2\pi\nu L_xL_y}\sum_{n=1}^{\infty}\bigg\{\sum_{m= 1}^{\infty}\bigg[ \sum_{j=1}^{3} \frac{1}{\lambda_{n,m}^{(j)} - i\delta\omega\tau_{A}} \\
&-&\frac{1}{\lambda_{n,m} - i\delta\omega\tau_{A}} \bigg] -\frac{2}{\lambda_{n,0} - i\delta\omega\tau_{A}}\bigg\}.
\end{eqnarray}
Let us analyze this expression in several limits. In a narrow wire case $L_y\ll L_x < L_A$, the contribution of the pseudospin singlet term $C^{(0)}_0$ 
dominates over the triplets $(L_y/L_x)\ln| L_A/L_y| \ll 1$
(similarly to the conductance fluctuations in graphene \cite{PhysRevB.77.193403}). 
Summing up over $n$ in the last term in Eq. \ref{correction_WL}, one obtains $\delta D/D  = -(\tau_A/2\pi\nu L_xL_y)[1- (\pi L_x/L_A)\mathrm{ctg}(\pi L_x/L_A) ]
$ at $\delta\omega=0$.
At the vicinity of the threshold, $L_x  =(1-\Delta)L_{A}$, it is enough to keep only single harmonics $\{n,m\}=\{1,0\}$, which gives
\begin{eqnarray}
\frac{\delta D}{D} =- \frac{\tau_A}{\pi \nu L_A L_y} \frac{1}{2\Delta}.
\end{eqnarray}
Correction to the diffusion coefficient has weak localization negative sign. 

In the wide contact case, $L_x \ll L_y < L_{A}$, both singlet and triplet components can equally contribute. Here the sign of the correction can change 
to positive. Indeed, at $\delta\omega=0$, one estimates 
\begin{eqnarray}\nonumber
\frac{\delta D}{D} &=&\frac{\tau_A}{4\nu L_a^2}\sum_{n\geq 1}\bigg\{  -\left[n^2 - \frac{L_x^2}{L_A^2}\right]^{-1/2} \\
&+&\sum_{j=1}^3 \left[n^2 - \frac{L_x^2}{L_A^2}\left(1- \frac{L_A^2}{D\pi^2 \tau_j^2}\right)\right]^{-1/2} \bigg\}.
\end{eqnarray}
Although, since the media is far from the threshold condition the logarithmic interference correction is not enhanced as compared to the
random media without amplification. 

It is also instructive to consider the case when the intervalley scattering processes at the boundaries are weak, so that 
all components of the Cooperon are given by expression in Eq. \ref{Result_F0} with $\tau_{A}^{-1} \rightarrow \tau_{A}^{-1} - \tau_j^{-1}$. In this situation, 
both the singlet and the triplet might contribute equally 
in the most singular zero-dimensional case, at $L_y\approx L_x=L_A(1-\Delta)$. For the lowest harmonic we obtain
\begin{eqnarray}\nonumber
\delta D/D &\propto&
-\frac{v^2\tau_{A}}{L_xL_y}\bigg\{ \frac{1}{2\Delta - i\delta\omega\tau_{A}} \\
&-& \sum_{j=1}^{3} \frac{1}{2\Delta + (\tau_j^{-1}- i\delta\omega)\tau_{A}} \bigg\}.
\end{eqnarray}
For example, at $\tau_1= \tau_2 \neq \tau_3$ and for zero frequency $\delta\omega=0$, with the decrease of $\Delta$ the sign of the interference correction
changes from positive to negative at $\Delta = [\sqrt{1+ 8(\tau_3/\tau_1)} -1]\tau_A/8\tau_3$ provided $\tau_A/\tau_j \ll 1$.

\section{Summary}\label{sec5}
In conclusion, we have explored the interplay of amplification and disorder on the propagation of light through a photonic honeycomb lattice. 
The honeycomb PhL possesses valley degree of freedom, which allows various scattering processes for light. 
The relative strength of the intervalley and intravalley scattering rates with respect to that of amplification background might
monitor the sign of the interference correction to light transmission and reflection coefficients. 

Finally, we have considered the term
$i\Gamma\sigma_{ab}^0$ describing amplification in the non-Hermitian Hamiltonian Eq. (\ref{Disorder_1}). One of the possible future directions for study is to
consider the case with loss-gain imbalance between the waveguides described by the term $i\vec{\gamma}\cdot\vec{\sigma}_{ab}$. 
It would be interesting to extend the research to this situation with strongly anisotropic light propagation and revisit the condition for the lasing threshold.

\section{acknowledgements}
We are thankful to A. Yu. Zyuzin for critical discussions. This work is supported by the Academy of Finland Grant No. 308339. 
A.A.Z. is grateful to the hospitality of the Pirinem School of Theoretical Physics. 
\bibliography{disorder_PhL_v2}

\begin{thebibliography}{49}%
\makeatletter
\providecommand \@ifxundefined [1]{%
 \@ifx{#1\undefined}
}%
\providecommand \@ifnum [1]{%
 \ifnum #1\expandafter \@firstoftwo
 \else \expandafter \@secondoftwo
 \fi
}%
\providecommand \@ifx [1]{%
 \ifx #1\expandafter \@firstoftwo
 \else \expandafter \@secondoftwo
 \fi
}%
\providecommand \natexlab [1]{#1}%
\providecommand \enquote  [1]{``#1''}%
\providecommand \bibnamefont  [1]{#1}%
\providecommand \bibfnamefont [1]{#1}%
\providecommand \citenamefont [1]{#1}%
\providecommand \href@noop [0]{\@secondoftwo}%
\providecommand \href [0]{\begingroup \@sanitize@url \@href}%
\providecommand \@href[1]{\@@startlink{#1}\@@href}%
\providecommand \@@href[1]{\endgroup#1\@@endlink}%
\providecommand \@sanitize@url [0]{\catcode `\\12\catcode `\$12\catcode
  `\&12\catcode `\#12\catcode `\^12\catcode `\_12\catcode `\%12\relax}%
\providecommand \@@startlink[1]{}%
\providecommand \@@endlink[0]{}%
\providecommand \url  [0]{\begingroup\@sanitize@url \@url }%
\providecommand \@url [1]{\endgroup\@href {#1}{\urlprefix }}%
\providecommand \urlprefix  [0]{URL }%
\providecommand \Eprint [0]{\href }%
\providecommand \doibase [0]{http://dx.doi.org/}%
\providecommand \selectlanguage [0]{\@gobble}%
\providecommand \bibinfo  [0]{\@secondoftwo}%
\providecommand \bibfield  [0]{\@secondoftwo}%
\providecommand \translation [1]{[#1]}%
\providecommand \BibitemOpen [0]{}%
\providecommand \bibitemStop [0]{}%
\providecommand \bibitemNoStop [0]{.\EOS\space}%
\providecommand \EOS [0]{\spacefactor3000\relax}%
\providecommand \BibitemShut  [1]{\csname bibitem#1\endcsname}%
\let\auto@bib@innerbib\@empty
\bibitem [{\citenamefont {Letokhov}(1968)}]{letokhov1968generation}%
  \BibitemOpen
  \bibfield  {author} {\bibinfo {author} {\bibfnamefont {V.~S.}\ \bibnamefont
  {Letokhov}},\ }\bibfield  {title} {\enquote {\bibinfo {title} {{Generation of
  light by a scattering medium with negative resonance absorption}},}\ }\href
  {http://www.jetp.ac.ru/cgi-bin/e/index/e/26/4/p835?a=list} {\bibfield
  {journal} {\bibinfo  {journal} {Sov. Phys. JETP}\ }\textbf {\bibinfo {volume}
  {26}},\ \bibinfo {pages} {835--840} (\bibinfo {year} {1968})}\BibitemShut
  {NoStop}%
\bibitem [{\citenamefont {Turitsyn}\ \emph {et~al.}(2010)\citenamefont
  {Turitsyn}, \citenamefont {Babin}, \citenamefont {El-Taher}, \citenamefont
  {Harper}, \citenamefont {Churkin}, \citenamefont {Kablukov}, \citenamefont
  {Ania-Castanon}, \citenamefont {Karalekas},\ and\ \citenamefont
  {Podivilov}}]{turitsyn2010random}%
  \BibitemOpen
  \bibfield  {author} {\bibinfo {author} {\bibfnamefont {S.~K.}\ \bibnamefont
  {Turitsyn}}, \bibinfo {author} {\bibfnamefont {S.~A.}\ \bibnamefont {Babin}},
  \bibinfo {author} {\bibfnamefont {A.~E.}\ \bibnamefont {El-Taher}}, \bibinfo
  {author} {\bibfnamefont {P.}~\bibnamefont {Harper}}, \bibinfo {author}
  {\bibfnamefont {D.~V.}\ \bibnamefont {Churkin}}, \bibinfo {author}
  {\bibfnamefont {S.~I.}\ \bibnamefont {Kablukov}}, \bibinfo {author}
  {\bibfnamefont {J.~D.}\ \bibnamefont {Ania-Castanon}}, \bibinfo {author}
  {\bibfnamefont {V.}~\bibnamefont {Karalekas}}, \ and\ \bibinfo {author}
  {\bibfnamefont {E.~V.}\ \bibnamefont {Podivilov}},\ }\bibfield  {title}
  {\enquote {\bibinfo {title} {Random distributed feedback fibre laser},}\
  }\href {\doibase https://doi.org/10.1038/nphoton.2010.4} {\bibfield
  {journal} {\bibinfo  {journal} {Nature Photonics}\ }\textbf {\bibinfo
  {volume} {4}},\ \bibinfo {pages} {231} (\bibinfo {year} {2010})}\BibitemShut
  {NoStop}%
\bibitem [{\citenamefont {Wiersma}(2013)}]{wiersma2013}%
  \BibitemOpen
  \bibfield  {author} {\bibinfo {author} {\bibfnamefont {D.~S}\ \bibnamefont
  {Wiersma}},\ }\bibfield  {title} {\enquote {\bibinfo {title} {Disordered
  photonics},}\ }\href {\doibase https://doi.org/10.1038/nphoton.2013.29}
  {\bibfield  {journal} {\bibinfo  {journal} {Nature Photonics}\ }\textbf
  {\bibinfo {volume} {7}},\ \bibinfo {pages} {188} (\bibinfo {year}
  {2013})}\BibitemShut {NoStop}%
\bibitem [{\citenamefont {Wiersma}(2008)}]{wiersma2008physics}%
  \BibitemOpen
  \bibfield  {author} {\bibinfo {author} {\bibfnamefont {D.~S}\ \bibnamefont
  {Wiersma}},\ }\bibfield  {title} {\enquote {\bibinfo {title} {The physics and
  applications of random lasers},}\ }\href {\doibase
  https://doi.org/10.1038/nphys971} {\bibfield  {journal} {\bibinfo  {journal}
  {Nature physics}\ }\textbf {\bibinfo {volume} {4}},\ \bibinfo {pages}
  {359--367} (\bibinfo {year} {2008})}\BibitemShut {NoStop}%
\bibitem [{\citenamefont {Cao}(2003)}]{cao2003lasing}%
  \BibitemOpen
  \bibfield  {author} {\bibinfo {author} {\bibfnamefont {H.}~\bibnamefont
  {Cao}},\ }\bibfield  {title} {\enquote {\bibinfo {title} {Lasing in random
  media},}\ }\href {\doibase https://doi.org/10.1088/0959-7174/13/3/201}
  {\bibfield  {journal} {\bibinfo  {journal} {Waves in random media}\ }\textbf
  {\bibinfo {volume} {13}},\ \bibinfo {pages} {R1--R39} (\bibinfo {year}
  {2003})}\BibitemShut {NoStop}%
\bibitem [{\citenamefont {Cao}\ \emph {et~al.}(2000)\citenamefont {Cao},
  \citenamefont {Xu}, \citenamefont {Zhang}, \citenamefont {Chang},
  \citenamefont {Ho}, \citenamefont {Seelig}, \citenamefont {Liu},\ and\
  \citenamefont {Chang}}]{PhysRevLett.84.5584}%
  \BibitemOpen
  \bibfield  {author} {\bibinfo {author} {\bibfnamefont {H.}~\bibnamefont
  {Cao}}, \bibinfo {author} {\bibfnamefont {J.~Y.}\ \bibnamefont {Xu}},
  \bibinfo {author} {\bibfnamefont {D.~Z.}\ \bibnamefont {Zhang}}, \bibinfo
  {author} {\bibfnamefont {S.-H.}\ \bibnamefont {Chang}}, \bibinfo {author}
  {\bibfnamefont {S.~T.}\ \bibnamefont {Ho}}, \bibinfo {author} {\bibfnamefont
  {E.~W.}\ \bibnamefont {Seelig}}, \bibinfo {author} {\bibfnamefont
  {X.}~\bibnamefont {Liu}}, \ and\ \bibinfo {author} {\bibfnamefont {R.~P.~H.}\
  \bibnamefont {Chang}},\ }\bibfield  {title} {\enquote {\bibinfo {title}
  {Spatial confinement of laser light in active random media},}\ }\href
  {\doibase 10.1103/PhysRevLett.84.5584} {\bibfield  {journal} {\bibinfo
  {journal} {Phys. Rev. Lett.}\ }\textbf {\bibinfo {volume} {84}},\ \bibinfo
  {pages} {5584--5587} (\bibinfo {year} {2000})}\BibitemShut {NoStop}%
\bibitem [{\citenamefont {Riboli}\ \emph {et~al.}(2014)\citenamefont {Riboli},
  \citenamefont {Caselli}, \citenamefont {Vignolini}, \citenamefont {Intonti},
  \citenamefont {Vynck}, \citenamefont {Barthelemy}, \citenamefont {Gerardino},
  \citenamefont {Balet}, \citenamefont {Li},\ and\ \citenamefont
  {Fiore}}]{riboli}%
  \BibitemOpen
  \bibfield  {author} {\bibinfo {author} {\bibfnamefont {F.}~\bibnamefont
  {Riboli}}, \bibinfo {author} {\bibfnamefont {N.}~\bibnamefont {Caselli}},
  \bibinfo {author} {\bibfnamefont {S.}~\bibnamefont {Vignolini}}, \bibinfo
  {author} {\bibfnamefont {F.}~\bibnamefont {Intonti}}, \bibinfo {author}
  {\bibfnamefont {K.}~\bibnamefont {Vynck}}, \bibinfo {author} {\bibfnamefont
  {P.}~\bibnamefont {Barthelemy}}, \bibinfo {author} {\bibfnamefont
  {A.}~\bibnamefont {Gerardino}}, \bibinfo {author} {\bibfnamefont
  {L.}~\bibnamefont {Balet}}, \bibinfo {author} {\bibfnamefont {L.~H.}\
  \bibnamefont {Li}}, \ and\ \bibinfo {author} {\bibfnamefont {A.}~\bibnamefont
  {Fiore}},\ }\bibfield  {title} {\enquote {\bibinfo {title} {Engineering of
  light confinement in strongly scattering disordered media},}\ }\href
  {\doibase https://doi.org/10.1038/nmat3966} {\bibfield  {journal} {\bibinfo
  {journal} {Nature materials}\ }\textbf {\bibinfo {volume} {13}},\ \bibinfo
  {pages} {720--725} (\bibinfo {year} {2014})}\BibitemShut {NoStop}%
\bibitem [{\citenamefont {Popoff}\ \emph {et~al.}(2014)\citenamefont {Popoff},
  \citenamefont {Goetschy}, \citenamefont {Liew}, \citenamefont {Stone},\ and\
  \citenamefont {Cao}}]{PhysRevLett.112.133903}%
  \BibitemOpen
  \bibfield  {author} {\bibinfo {author} {\bibfnamefont {S.~M.}\ \bibnamefont
  {Popoff}}, \bibinfo {author} {\bibfnamefont {A.}~\bibnamefont {Goetschy}},
  \bibinfo {author} {\bibfnamefont {S.~F.}\ \bibnamefont {Liew}}, \bibinfo
  {author} {\bibfnamefont {A.~D.}\ \bibnamefont {Stone}}, \ and\ \bibinfo
  {author} {\bibfnamefont {H.}~\bibnamefont {Cao}},\ }\bibfield  {title}
  {\enquote {\bibinfo {title} {Coherent control of total transmission of light
  through disordered media},}\ }\href {\doibase 10.1103/PhysRevLett.112.133903}
  {\bibfield  {journal} {\bibinfo  {journal} {Phys. Rev. Lett.}\ }\textbf
  {\bibinfo {volume} {112}},\ \bibinfo {pages} {133903} (\bibinfo {year}
  {2014})}\BibitemShut {NoStop}%
\bibitem [{\citenamefont {Starshynov}\ \emph {et~al.}(2019)\citenamefont
  {Starshynov}, \citenamefont {Ghafur}, \citenamefont {Fitches},\ and\
  \citenamefont {Faccio}}]{PhysRevApplied.12.064045}%
  \BibitemOpen
  \bibfield  {author} {\bibinfo {author} {\bibfnamefont {I.}~\bibnamefont
  {Starshynov}}, \bibinfo {author} {\bibfnamefont {O.}~\bibnamefont {Ghafur}},
  \bibinfo {author} {\bibfnamefont {J.}~\bibnamefont {Fitches}}, \ and\
  \bibinfo {author} {\bibfnamefont {D.}~\bibnamefont {Faccio}},\ }\bibfield
  {title} {\enquote {\bibinfo {title} {Coherent control of light for
  non-line-of-sight imaging},}\ }\href {\doibase
  10.1103/PhysRevApplied.12.064045} {\bibfield  {journal} {\bibinfo  {journal}
  {Phys. Rev. Applied}\ }\textbf {\bibinfo {volume} {12}},\ \bibinfo {pages}
  {064045} (\bibinfo {year} {2019})}\BibitemShut {NoStop}%
\bibitem [{\citenamefont {Yun}\ and\ \citenamefont
  {Kwok}(2017)}]{yun2017light}%
  \BibitemOpen
  \bibfield  {author} {\bibinfo {author} {\bibfnamefont {S.~H.}\ \bibnamefont
  {Yun}}\ and\ \bibinfo {author} {\bibfnamefont {S.~J.~J.}\ \bibnamefont
  {Kwok}},\ }\bibfield  {title} {\enquote {\bibinfo {title} {Light in
  diagnosis, therapy and surgery},}\ }\href {\doibase
  https://doi.org/10.1038/s41551-016-0008} {\bibfield  {journal} {\bibinfo
  {journal} {Nature biomedical engineering}\ }\textbf {\bibinfo {volume} {1}},\
  \bibinfo {pages} {1--16} (\bibinfo {year} {2017})}\BibitemShut {NoStop}%
\bibitem [{\citenamefont {Polson}\ and\ \citenamefont
  {Vardeny}(2004)}]{polson2004random}%
  \BibitemOpen
  \bibfield  {author} {\bibinfo {author} {\bibfnamefont {R.~C.}\ \bibnamefont
  {Polson}}\ and\ \bibinfo {author} {\bibfnamefont {Z.~V.}\ \bibnamefont
  {Vardeny}},\ }\bibfield  {title} {\enquote {\bibinfo {title} {Random lasing
  in human tissues},}\ }\href {\doibase https://doi.org/10.1063/1.1782259}
  {\bibfield  {journal} {\bibinfo  {journal} {Applied physics letters}\
  }\textbf {\bibinfo {volume} {85}},\ \bibinfo {pages} {1289--1291} (\bibinfo
  {year} {2004})}\BibitemShut {NoStop}%
\bibitem [{\citenamefont {Watson}(1969)}]{doi:10.1063}%
  \BibitemOpen
  \bibfield  {author} {\bibinfo {author} {\bibfnamefont {K.~M.}\ \bibnamefont
  {Watson}},\ }\bibfield  {title} {\enquote {\bibinfo {title} {{Multiple
  Scattering of Electromagnetic Waves in an Underdense Plasma}},}\ }\href
  {\doibase 10.1063/1.1664895} {\bibfield  {journal} {\bibinfo  {journal} {J.
  Math. Phys.}\ }\textbf {\bibinfo {volume} {10}},\ \bibinfo {pages} {688--702}
  (\bibinfo {year} {1969})}\BibitemShut {NoStop}%
\bibitem [{\citenamefont {Wiersma}\ \emph {et~al.}(1997)\citenamefont
  {Wiersma}, \citenamefont {Bartolini}, \citenamefont {Lagendijk},\ and\
  \citenamefont {Righini}}]{wiersma1997}%
  \BibitemOpen
  \bibfield  {author} {\bibinfo {author} {\bibfnamefont {D.~S.}\ \bibnamefont
  {Wiersma}}, \bibinfo {author} {\bibfnamefont {P.}~\bibnamefont {Bartolini}},
  \bibinfo {author} {\bibfnamefont {A.}~\bibnamefont {Lagendijk}}, \ and\
  \bibinfo {author} {\bibfnamefont {R.}~\bibnamefont {Righini}},\ }\bibfield
  {title} {\enquote {\bibinfo {title} {{Localization of light in a disordered
  medium}},}\ }\href {\doibase 10.1038/37757} {\bibfield  {journal} {\bibinfo
  {journal} {Nature}\ }\textbf {\bibinfo {volume} {390}},\ \bibinfo {pages}
  {671--673} (\bibinfo {year} {1997})}\BibitemShut {NoStop}%
\bibitem [{\citenamefont {Burkov}\ and\ \citenamefont
  {Zyuzin}(1997)}]{PhysRevB.55.5736}%
  \BibitemOpen
  \bibfield  {author} {\bibinfo {author} {\bibfnamefont {A.~A.}\ \bibnamefont
  {Burkov}}\ and\ \bibinfo {author} {\bibfnamefont {A.~Yu.}\ \bibnamefont
  {Zyuzin}},\ }\bibfield  {title} {\enquote {\bibinfo {title} {{Correlations in
  transmission of light through a disordered amplifying medium}},}\ }\href
  {\doibase 10.1103/PhysRevB.55.5736} {\bibfield  {journal} {\bibinfo
  {journal} {Phys. Rev. B}\ }\textbf {\bibinfo {volume} {55}},\ \bibinfo
  {pages} {5736--5741} (\bibinfo {year} {1997})}\BibitemShut {NoStop}%
\bibitem [{\citenamefont {Stephen}(1991)}]{Stephen}%
  \BibitemOpen
  \bibfield  {author} {\bibinfo {author} {\bibfnamefont {M.~J.}\ \bibnamefont
  {Stephen}},\ }\bibfield  {title} {\enquote {\bibinfo {title} {{Interference,
  fluctuations and correlations in the diffusive scattering of light from
  disordered medium}},}\ }in\ \href@noop {} {\emph {\bibinfo {booktitle}
  {{Mesoscopic Phenomena in Solids}}}},\ \bibinfo {editor} {edited by\ \bibinfo
  {editor} {\bibfnamefont {B.~L.}\ \bibnamefont {Altshuler}}, \bibinfo {editor}
  {\bibfnamefont {P.~A.}\ \bibnamefont {Lee}}, \ and\ \bibinfo {editor}
  {\bibfnamefont {W.~R.}\ \bibnamefont {Webb}}}\ (\bibinfo  {publisher}
  {Elsevier Science Publishers, Amsterdam},\ \bibinfo {year}
  {1991})\BibitemShut {NoStop}%
\bibitem [{\citenamefont {Zyuzin}(1995)}]{PhysRevE.51.5274}%
  \BibitemOpen
  \bibfield  {author} {\bibinfo {author} {\bibfnamefont {A.~Yu.}\ \bibnamefont
  {Zyuzin}},\ }\bibfield  {title} {\enquote {\bibinfo {title} {{Transmission
  fluctuations and spectral rigidity of lasing states in a random amplifying
  medium}},}\ }\href {\doibase 10.1103/PhysRevE.51.5274} {\bibfield  {journal}
  {\bibinfo  {journal} {Phys. Rev. E}\ }\textbf {\bibinfo {volume} {51}},\
  \bibinfo {pages} {5274--5278} (\bibinfo {year} {1995})}\BibitemShut {NoStop}%
\bibitem [{\citenamefont {Joshi}\ and\ \citenamefont
  {Jayannavar}(1997)}]{PhysRevB.56.12038}%
  \BibitemOpen
  \bibfield  {author} {\bibinfo {author} {\bibfnamefont {S.~K.}\ \bibnamefont
  {Joshi}}\ and\ \bibinfo {author} {\bibfnamefont {A.~M.}\ \bibnamefont
  {Jayannavar}},\ }\bibfield  {title} {\enquote {\bibinfo {title} {Transmission
  and reflection from a disordered lasing medium},}\ }\href {\doibase
  10.1103/PhysRevB.56.12038} {\bibfield  {journal} {\bibinfo  {journal} {Phys.
  Rev. B}\ }\textbf {\bibinfo {volume} {56}},\ \bibinfo {pages} {12038--12041}
  (\bibinfo {year} {1997})}\BibitemShut {NoStop}%
\bibitem [{\citenamefont {Zyuzin}(1994)}]{Zyuzin_1994}%
  \BibitemOpen
  \bibfield  {author} {\bibinfo {author} {\bibfnamefont {A.~Yu.}\ \bibnamefont
  {Zyuzin}},\ }\bibfield  {title} {\enquote {\bibinfo {title} {Weak
  localization in backscattering from an amplifying medium},}\ }\href {\doibase
  10.1209/0295-5075/26/7/007} {\bibfield  {journal} {\bibinfo  {journal}
  {Europhys. Lett.}\ }\textbf {\bibinfo {volume} {26}},\ \bibinfo {pages}
  {517--520} (\bibinfo {year} {1994})}\BibitemShut {NoStop}%
\bibitem [{\citenamefont {Liew}\ \emph {et~al.}(2014)\citenamefont {Liew},
  \citenamefont {Popoff}, \citenamefont {Mosk}, \citenamefont {Vos},\ and\
  \citenamefont {Cao}}]{PhysRevB.89.224202}%
  \BibitemOpen
  \bibfield  {author} {\bibinfo {author} {\bibfnamefont {S.~F.}\ \bibnamefont
  {Liew}}, \bibinfo {author} {\bibfnamefont {S.~M.}\ \bibnamefont {Popoff}},
  \bibinfo {author} {\bibfnamefont {A.~P.}\ \bibnamefont {Mosk}}, \bibinfo
  {author} {\bibfnamefont {W.~L.}\ \bibnamefont {Vos}}, \ and\ \bibinfo
  {author} {\bibfnamefont {H.}~\bibnamefont {Cao}},\ }\bibfield  {title}
  {\enquote {\bibinfo {title} {{Transmission channels for light in absorbing
  random media: From diffusive to ballistic-like transport}},}\ }\href
  {\doibase 10.1103/PhysRevB.89.224202} {\bibfield  {journal} {\bibinfo
  {journal} {Phys. Rev. B}\ }\textbf {\bibinfo {volume} {89}},\ \bibinfo
  {pages} {224202} (\bibinfo {year} {2014})}\BibitemShut {NoStop}%
\bibitem [{\citenamefont {Wiersma}\ and\ \citenamefont
  {Lagendijk}(1996)}]{PhysRevE.54.4256}%
  \BibitemOpen
  \bibfield  {author} {\bibinfo {author} {\bibfnamefont {D.~S.}\ \bibnamefont
  {Wiersma}}\ and\ \bibinfo {author} {\bibfnamefont {A.}~\bibnamefont
  {Lagendijk}},\ }\bibfield  {title} {\enquote {\bibinfo {title} {Light
  diffusion with gain and random lasers},}\ }\href {\doibase
  10.1103/PhysRevE.54.4256} {\bibfield  {journal} {\bibinfo  {journal} {Phys.
  Rev. E}\ }\textbf {\bibinfo {volume} {54}},\ \bibinfo {pages} {4256--4265}
  (\bibinfo {year} {1996})}\BibitemShut {NoStop}%
\bibitem [{\citenamefont {Zhang}(1995)}]{PhysRevB.52.7960}%
  \BibitemOpen
  \bibfield  {author} {\bibinfo {author} {\bibfnamefont {Z.-Q.}\ \bibnamefont
  {Zhang}},\ }\bibfield  {title} {\enquote {\bibinfo {title} {{Light
  amplification and localization in randomly layered media with gain}},}\
  }\href {\doibase 10.1103/PhysRevB.52.7960} {\bibfield  {journal} {\bibinfo
  {journal} {Phys. Rev. B}\ }\textbf {\bibinfo {volume} {52}},\ \bibinfo
  {pages} {7960--7964} (\bibinfo {year} {1995})}\BibitemShut {NoStop}%
\bibitem [{\citenamefont {Pradhan}\ and\ \citenamefont
  {Kumar}(1994)}]{PhysRevB.50.9644}%
  \BibitemOpen
  \bibfield  {author} {\bibinfo {author} {\bibfnamefont {P.}~\bibnamefont
  {Pradhan}}\ and\ \bibinfo {author} {\bibfnamefont {N.}~\bibnamefont
  {Kumar}},\ }\bibfield  {title} {\enquote {\bibinfo {title} {Localization of
  light in coherently amplifying random media},}\ }\href {\doibase
  10.1103/PhysRevB.50.9644} {\bibfield  {journal} {\bibinfo  {journal} {Phys.
  Rev. B}\ }\textbf {\bibinfo {volume} {50}},\ \bibinfo {pages} {9644--9647}
  (\bibinfo {year} {1994})}\BibitemShut {NoStop}%
\bibitem [{\citenamefont {de~Oliveira}\ \emph {et~al.}(1997)\citenamefont
  {de~Oliveira}, \citenamefont {McGreevy},\ and\ \citenamefont
  {Lawandy}}]{deOliveira:97}%
  \BibitemOpen
  \bibfield  {author} {\bibinfo {author} {\bibfnamefont {P.~C.}\ \bibnamefont
  {de~Oliveira}}, \bibinfo {author} {\bibfnamefont {J.~A.}\ \bibnamefont
  {McGreevy}}, \ and\ \bibinfo {author} {\bibfnamefont {N.~M.}\ \bibnamefont
  {Lawandy}},\ }\bibfield  {title} {\enquote {\bibinfo {title}
  {External-feedback effects in high-gain scattering media},}\ }\href {\doibase
  10.1364/OL.22.000895} {\bibfield  {journal} {\bibinfo  {journal} {Opt.
  Lett.}\ }\textbf {\bibinfo {volume} {22}},\ \bibinfo {pages} {895--897}
  (\bibinfo {year} {1997})}\BibitemShut {NoStop}%
\bibitem [{\citenamefont {de~Oliveira}\ \emph {et~al.}(1996)\citenamefont
  {de~Oliveira}, \citenamefont {Perkins},\ and\ \citenamefont
  {Lawandy}}]{deOliveira:96}%
  \BibitemOpen
  \bibfield  {author} {\bibinfo {author} {\bibfnamefont {P.~C.}\ \bibnamefont
  {de~Oliveira}}, \bibinfo {author} {\bibfnamefont {A.~E.}\ \bibnamefont
  {Perkins}}, \ and\ \bibinfo {author} {\bibfnamefont {N.~M.}\ \bibnamefont
  {Lawandy}},\ }\bibfield  {title} {\enquote {\bibinfo {title} {Coherent
  backscattering from high-gain scattering media},}\ }\href {\doibase
  10.1364/OL.21.001685} {\bibfield  {journal} {\bibinfo  {journal} {Opt.
  Lett.}\ }\textbf {\bibinfo {volume} {21}},\ \bibinfo {pages} {1685--1687}
  (\bibinfo {year} {1996})}\BibitemShut {NoStop}%
\bibitem [{\citenamefont {Ling}\ \emph {et~al.}(2001)\citenamefont {Ling},
  \citenamefont {Cao}, \citenamefont {Burin}, \citenamefont {Ratner},
  \citenamefont {Liu},\ and\ \citenamefont {Chang}}]{PhysRevA.64.063808}%
  \BibitemOpen
  \bibfield  {author} {\bibinfo {author} {\bibfnamefont {Y.}~\bibnamefont
  {Ling}}, \bibinfo {author} {\bibfnamefont {H.}~\bibnamefont {Cao}}, \bibinfo
  {author} {\bibfnamefont {A.~L.}\ \bibnamefont {Burin}}, \bibinfo {author}
  {\bibfnamefont {M.~A.}\ \bibnamefont {Ratner}}, \bibinfo {author}
  {\bibfnamefont {X.}~\bibnamefont {Liu}}, \ and\ \bibinfo {author}
  {\bibfnamefont {R.~P.~H.}\ \bibnamefont {Chang}},\ }\bibfield  {title}
  {\enquote {\bibinfo {title} {Investigation of random lasers with resonant
  feedback},}\ }\href {\doibase 10.1103/PhysRevA.64.063808} {\bibfield
  {journal} {\bibinfo  {journal} {Phys. Rev. A}\ }\textbf {\bibinfo {volume}
  {64}},\ \bibinfo {pages} {063808} (\bibinfo {year} {2001})}\BibitemShut
  {NoStop}%
\bibitem [{\citenamefont {Gottardo}\ \emph {et~al.}(2004)\citenamefont
  {Gottardo}, \citenamefont {Cavalieri}, \citenamefont {Yaroshchuk},\ and\
  \citenamefont {Wiersma}}]{PhysRevLett.93.263901}%
  \BibitemOpen
  \bibfield  {author} {\bibinfo {author} {\bibfnamefont {S.}~\bibnamefont
  {Gottardo}}, \bibinfo {author} {\bibfnamefont {S.}~\bibnamefont {Cavalieri}},
  \bibinfo {author} {\bibfnamefont {O.}~\bibnamefont {Yaroshchuk}}, \ and\
  \bibinfo {author} {\bibfnamefont {D.~S.}\ \bibnamefont {Wiersma}},\
  }\bibfield  {title} {\enquote {\bibinfo {title} {Quasi-two-dimensional
  diffusive random laser action},}\ }\href {\doibase
  10.1103/PhysRevLett.93.263901} {\bibfield  {journal} {\bibinfo  {journal}
  {Phys. Rev. Lett.}\ }\textbf {\bibinfo {volume} {93}},\ \bibinfo {pages}
  {263901} (\bibinfo {year} {2004})}\BibitemShut {NoStop}%
\bibitem [{\citenamefont {Wiersma}\ \emph {et~al.}(1995)\citenamefont
  {Wiersma}, \citenamefont {van Albada},\ and\ \citenamefont
  {Lagendijk}}]{PhysRevLett.75.1739}%
  \BibitemOpen
  \bibfield  {author} {\bibinfo {author} {\bibfnamefont {D.~S.}\ \bibnamefont
  {Wiersma}}, \bibinfo {author} {\bibfnamefont {M.~P.}\ \bibnamefont {van
  Albada}}, \ and\ \bibinfo {author} {\bibfnamefont {A.}~\bibnamefont
  {Lagendijk}},\ }\bibfield  {title} {\enquote {\bibinfo {title} {Coherent
  backscattering of light from amplifying random media},}\ }\href {\doibase
  10.1103/PhysRevLett.75.1739} {\bibfield  {journal} {\bibinfo  {journal}
  {Phys. Rev. Lett.}\ }\textbf {\bibinfo {volume} {75}},\ \bibinfo {pages}
  {1739--1742} (\bibinfo {year} {1995})}\BibitemShut {NoStop}%
\bibitem [{\citenamefont {Volovik}(2003)}]{volovik_book}%
  \BibitemOpen
  \bibfield  {author} {\bibinfo {author} {\bibfnamefont {G.~E.}\ \bibnamefont
  {Volovik}},\ }\href@noop {} {\emph {\bibinfo {title} {{The Universe in a
  Helium Droplet}}}}\ (\bibinfo  {publisher} {Clarendon Press, Oxford},\
  \bibinfo {year} {2003})\BibitemShut {NoStop}%
\bibitem [{\citenamefont {Hasan}\ and\ \citenamefont
  {Kane}(2010)}]{RevModPhys.82.3045}%
  \BibitemOpen
  \bibfield  {author} {\bibinfo {author} {\bibfnamefont {M.~Z.}\ \bibnamefont
  {Hasan}}\ and\ \bibinfo {author} {\bibfnamefont {C.~L.}\ \bibnamefont
  {Kane}},\ }\bibfield  {title} {\enquote {\bibinfo {title} {Colloquium:
  Topological insulators},}\ }\href {\doibase 10.1103/RevModPhys.82.3045}
  {\bibfield  {journal} {\bibinfo  {journal} {Rev. Mod. Phys.}\ }\textbf
  {\bibinfo {volume} {82}},\ \bibinfo {pages} {3045--3067} (\bibinfo {year}
  {2010})}\BibitemShut {NoStop}%
\bibitem [{\citenamefont {Haldane}\ and\ \citenamefont
  {Raghu}(2008)}]{PhysRevLett.100.013904}%
  \BibitemOpen
  \bibfield  {author} {\bibinfo {author} {\bibfnamefont {F.~D.~M.}\
  \bibnamefont {Haldane}}\ and\ \bibinfo {author} {\bibfnamefont
  {S.}~\bibnamefont {Raghu}},\ }\bibfield  {title} {\enquote {\bibinfo {title}
  {{Possible realization of directional optical waveguides in photonic crystals
  with broken time-reversal symmetry}},}\ }\href {\doibase
  10.1103/PhysRevLett.100.013904} {\bibfield  {journal} {\bibinfo  {journal}
  {Phys. Rev. Lett.}\ }\textbf {\bibinfo {volume} {100}},\ \bibinfo {pages}
  {013904} (\bibinfo {year} {2008})}\BibitemShut {NoStop}%
\bibitem [{\citenamefont {Plihal}\ and\ \citenamefont
  {Maradudin}(1991)}]{PhysRevB.44.8565}%
  \BibitemOpen
  \bibfield  {author} {\bibinfo {author} {\bibfnamefont {M.}~\bibnamefont
  {Plihal}}\ and\ \bibinfo {author} {\bibfnamefont {A.~A.}\ \bibnamefont
  {Maradudin}},\ }\bibfield  {title} {\enquote {\bibinfo {title} {{Photonic
  band structure of two-dimensional systems: The triangular lattice}},}\ }\href
  {\doibase 10.1103/PhysRevB.44.8565} {\bibfield  {journal} {\bibinfo
  {journal} {Phys. Rev. B}\ }\textbf {\bibinfo {volume} {44}},\ \bibinfo
  {pages} {8565--8571} (\bibinfo {year} {1991})}\BibitemShut {NoStop}%
\bibitem [{\citenamefont {Bender}(2007)}]{Bender_sense}%
  \BibitemOpen
  \bibfield  {author} {\bibinfo {author} {\bibfnamefont {C.~M.}\ \bibnamefont
  {Bender}},\ }\bibfield  {title} {\enquote {\bibinfo {title} {{Making sense of
  non-Hermitian Hamiltonians}},}\ }\href {\doibase 10.1088/0034-4885/70/6/R03}
  {\bibfield  {journal} {\bibinfo  {journal} {Rep. Prog. Phys.}\ }\textbf
  {\bibinfo {volume} {70}},\ \bibinfo {pages} {947} (\bibinfo {year}
  {2007})}\BibitemShut {NoStop}%
\bibitem [{\citenamefont {Berry}(2004)}]{Berry}%
  \BibitemOpen
  \bibfield  {author} {\bibinfo {author} {\bibfnamefont {M.}~\bibnamefont
  {Berry}},\ }\bibfield  {title} {\enquote {\bibinfo {title} {{Physics of
  Nonhermitian Degeneracies}},}\ }\href {\doibase
  10.1023/B:CJOP.0000044002.05657.04} {\bibfield  {journal} {\bibinfo
  {journal} {Czechoslov. J. Phys.}\ }\textbf {\bibinfo {volume} {54}},\
  \bibinfo {pages} {1039} (\bibinfo {year} {2004})}\BibitemShut {NoStop}%
\bibitem [{\citenamefont {Ozawa}\ \emph {et~al.}(2019)\citenamefont {Ozawa},
  \citenamefont {Price}, \citenamefont {Amo}, \citenamefont {Goldman},
  \citenamefont {Hafezi}, \citenamefont {Lu}, \citenamefont {Rechtsman},
  \citenamefont {Schuster}, \citenamefont {Simon}, \citenamefont {Zilberberg},\
  and\ \citenamefont {Carusotto}}]{Review_photonics_2018}%
  \BibitemOpen
  \bibfield  {author} {\bibinfo {author} {\bibfnamefont {T.}~\bibnamefont
  {Ozawa}}, \bibinfo {author} {\bibfnamefont {H.~M.}\ \bibnamefont {Price}},
  \bibinfo {author} {\bibfnamefont {A.}~\bibnamefont {Amo}}, \bibinfo {author}
  {\bibfnamefont {N.}~\bibnamefont {Goldman}}, \bibinfo {author} {\bibfnamefont
  {M.}~\bibnamefont {Hafezi}}, \bibinfo {author} {\bibfnamefont
  {L.}~\bibnamefont {Lu}}, \bibinfo {author} {\bibfnamefont {M.~C.}\
  \bibnamefont {Rechtsman}}, \bibinfo {author} {\bibfnamefont {D.}~\bibnamefont
  {Schuster}}, \bibinfo {author} {\bibfnamefont {J.}~\bibnamefont {Simon}},
  \bibinfo {author} {\bibfnamefont {O.}~\bibnamefont {Zilberberg}}, \ and\
  \bibinfo {author} {\bibfnamefont {I.}~\bibnamefont {Carusotto}},\ }\bibfield
  {title} {\enquote {\bibinfo {title} {{Topological photonics}},}\ }\href
  {\doibase 10.1103/RevModPhys.91.015006} {\bibfield  {journal} {\bibinfo
  {journal} {Rev. Mod. Phys.}\ }\textbf {\bibinfo {volume} {91}},\ \bibinfo
  {pages} {015006} (\bibinfo {year} {2019})}\BibitemShut {NoStop}%
\bibitem [{\citenamefont {Khanikaev}\ \emph {et~al.}(2012)\citenamefont
  {Khanikaev}, \citenamefont {Mousavi}, \citenamefont {Tse}, \citenamefont
  {Kargarian}, \citenamefont {MacDonald},\ and\ \citenamefont
  {Shvets}}]{Alexander}%
  \BibitemOpen
  \bibfield  {author} {\bibinfo {author} {\bibfnamefont {B.~A.}\ \bibnamefont
  {Khanikaev}}, \bibinfo {author} {\bibfnamefont {S.~H.}\ \bibnamefont
  {Mousavi}}, \bibinfo {author} {\bibfnamefont {W-K.}\ \bibnamefont {Tse}},
  \bibinfo {author} {\bibfnamefont {M.}~\bibnamefont {Kargarian}}, \bibinfo
  {author} {\bibfnamefont {A.~H.}\ \bibnamefont {MacDonald}}, \ and\ \bibinfo
  {author} {\bibfnamefont {G.}~\bibnamefont {Shvets}},\ }\bibfield  {title}
  {\enquote {\bibinfo {title} {Photonic topological insulators},}\ }\href
  {\doibase https://doi.org/10.1038/nmat3520} {\bibfield  {journal} {\bibinfo
  {journal} {Nature Materials}\ }\textbf {\bibinfo {volume} {12}},\ \bibinfo
  {pages} {233--239} (\bibinfo {year} {2012})}\BibitemShut {NoStop}%
\bibitem [{\citenamefont {Konotop}\ \emph {et~al.}(2016)\citenamefont
  {Konotop}, \citenamefont {Yang},\ and\ \citenamefont
  {Zezyulin}}]{Konotop_RevModPhys}%
  \BibitemOpen
  \bibfield  {author} {\bibinfo {author} {\bibfnamefont {V.~V.}\ \bibnamefont
  {Konotop}}, \bibinfo {author} {\bibfnamefont {J.}~\bibnamefont {Yang}}, \
  and\ \bibinfo {author} {\bibfnamefont {D.~A.}\ \bibnamefont {Zezyulin}},\
  }\bibfield  {title} {\enquote {\bibinfo {title} {{Nonlinear waves in
  $\mathcal{PT}$-symmetric systems}},}\ }\href {\doibase
  10.1103/RevModPhys.88.035002} {\bibfield  {journal} {\bibinfo  {journal}
  {Rev. Mod. Phys.}\ }\textbf {\bibinfo {volume} {88}},\ \bibinfo {pages}
  {035002} (\bibinfo {year} {2016})}\BibitemShut {NoStop}%
\bibitem [{\citenamefont {Sepkhanov}\ \emph {et~al.}(2007)\citenamefont
  {Sepkhanov}, \citenamefont {Bazaliy},\ and\ \citenamefont
  {Beenakker}}]{PhysRevA.75.063813}%
  \BibitemOpen
  \bibfield  {author} {\bibinfo {author} {\bibfnamefont {R.~A.}\ \bibnamefont
  {Sepkhanov}}, \bibinfo {author} {\bibfnamefont {Ya.~B.}\ \bibnamefont
  {Bazaliy}}, \ and\ \bibinfo {author} {\bibfnamefont {C.~W.~J.}\ \bibnamefont
  {Beenakker}},\ }\bibfield  {title} {\enquote {\bibinfo {title} {{Extremal
  transmission at the Dirac point of a photonic band structure}},}\ }\href
  {\doibase 10.1103/PhysRevA.75.063813} {\bibfield  {journal} {\bibinfo
  {journal} {Phys. Rev. A}\ }\textbf {\bibinfo {volume} {75}},\ \bibinfo
  {pages} {063813} (\bibinfo {year} {2007})}\BibitemShut {NoStop}%
\bibitem [{\citenamefont {Wang}\ \emph {et~al.}(2014)\citenamefont {Wang},
  \citenamefont {Jiang}, \citenamefont {Yan}, \citenamefont {Deng},
  \citenamefont {Sun}, \citenamefont {Li}, \citenamefont {Shi},\ and\
  \citenamefont {Chen}}]{Wang_2014}%
  \BibitemOpen
  \bibfield  {author} {\bibinfo {author} {\bibfnamefont {X.}~\bibnamefont
  {Wang}}, \bibinfo {author} {\bibfnamefont {H.~T.}\ \bibnamefont {Jiang}},
  \bibinfo {author} {\bibfnamefont {C.}~\bibnamefont {Yan}}, \bibinfo {author}
  {\bibfnamefont {F.~S.}\ \bibnamefont {Deng}}, \bibinfo {author}
  {\bibfnamefont {Y.}~\bibnamefont {Sun}}, \bibinfo {author} {\bibfnamefont
  {Y.~H.}\ \bibnamefont {Li}}, \bibinfo {author} {\bibfnamefont {Y.~L.}\
  \bibnamefont {Shi}}, \ and\ \bibinfo {author} {\bibfnamefont
  {H.}~\bibnamefont {Chen}},\ }\bibfield  {title} {\enquote {\bibinfo {title}
  {{Transmission properties near Dirac-like point in two-dimensional dielectric
  photonic crystals}},}\ }\href {\doibase 10.1209/0295-5075/108/14002}
  {\bibfield  {journal} {\bibinfo  {journal} {Europhys. Lett.}\ }\textbf
  {\bibinfo {volume} {108}},\ \bibinfo {pages} {14002} (\bibinfo {year}
  {2014})}\BibitemShut {NoStop}%
\bibitem [{\citenamefont {Sepkhanov}\ \emph {et~al.}(2009)\citenamefont
  {Sepkhanov}, \citenamefont {Ossipov},\ and\ \citenamefont
  {Beenakker}}]{Sepkhanov_2009}%
  \BibitemOpen
  \bibfield  {author} {\bibinfo {author} {\bibfnamefont {R.~A.}\ \bibnamefont
  {Sepkhanov}}, \bibinfo {author} {\bibfnamefont {A.}~\bibnamefont {Ossipov}},
  \ and\ \bibinfo {author} {\bibfnamefont {C.~W.~J.}\ \bibnamefont
  {Beenakker}},\ }\bibfield  {title} {\enquote {\bibinfo {title} {{Extinction
  of coherent backscattering by a disordered photonic crystal with a Dirac
  spectrum}},}\ }\href {\doibase 10.1209/0295-5075/85/14005} {\bibfield
  {journal} {\bibinfo  {journal} {Europhys. Lett.}\ }\textbf {\bibinfo {volume}
  {85}},\ \bibinfo {pages} {14005} (\bibinfo {year} {2009})}\BibitemShut
  {NoStop}%
\bibitem [{\citenamefont {Ando}\ and\ \citenamefont
  {Nakanishi}(1998)}]{Ando_Nakanishi}%
  \BibitemOpen
  \bibfield  {author} {\bibinfo {author} {\bibfnamefont {T.}~\bibnamefont
  {Ando}}\ and\ \bibinfo {author} {\bibfnamefont {T.}~\bibnamefont
  {Nakanishi}},\ }\bibfield  {title} {\enquote {\bibinfo {title} {{Impurity
  Scattering in Carbon Nanotubes - Absence of Back Scattering}},}\ }\href
  {\doibase 10.1143/JPSJ.67.1704} {\bibfield  {journal} {\bibinfo  {journal}
  {J. Phys. Soc. Jpn}\ }\textbf {\bibinfo {volume} {67}},\ \bibinfo {pages}
  {1704--1713} (\bibinfo {year} {1998})}\BibitemShut {NoStop}%
\bibitem [{\citenamefont {Suzuura}\ and\ \citenamefont
  {Ando}(2002)}]{Suzuura_Ando}%
  \BibitemOpen
  \bibfield  {author} {\bibinfo {author} {\bibfnamefont {H.}~\bibnamefont
  {Suzuura}}\ and\ \bibinfo {author} {\bibfnamefont {T.}~\bibnamefont {Ando}},\
  }\bibfield  {title} {\enquote {\bibinfo {title} {{Crossover from symplectic
  to orthogonal class in a two-dimensional honeycomb lattice}},}\ }\href
  {\doibase 10.1103/PhysRevLett.89.266603} {\bibfield  {journal} {\bibinfo
  {journal} {Phys. Rev. Lett.}\ }\textbf {\bibinfo {volume} {89}},\ \bibinfo
  {pages} {266603} (\bibinfo {year} {2002})}\BibitemShut {NoStop}%
\bibitem [{\citenamefont {Ni}\ \emph {et~al.}(2018)\citenamefont {Ni},
  \citenamefont {Smirnova}, \citenamefont {Poddubny}, \citenamefont {Leykam},
  \citenamefont {Chong},\ and\ \citenamefont {Khanikaev}}]{PhysRevB.98.165129}%
  \BibitemOpen
  \bibfield  {author} {\bibinfo {author} {\bibfnamefont {X.}~\bibnamefont
  {Ni}}, \bibinfo {author} {\bibfnamefont {D.}~\bibnamefont {Smirnova}},
  \bibinfo {author} {\bibfnamefont {A.}~\bibnamefont {Poddubny}}, \bibinfo
  {author} {\bibfnamefont {D.}~\bibnamefont {Leykam}}, \bibinfo {author}
  {\bibfnamefont {Y.}~\bibnamefont {Chong}}, \ and\ \bibinfo {author}
  {\bibfnamefont {A.~B.}\ \bibnamefont {Khanikaev}},\ }\bibfield  {title}
  {\enquote {\bibinfo {title} {{$\mathcal{PT}$ phase transitions of edge states
  at $\mathcal{PT}$ symmetric interfaces in non-Hermitian topological
  insulators}},}\ }\href {\doibase 10.1103/PhysRevB.98.165129} {\bibfield
  {journal} {\bibinfo  {journal} {Phys. Rev. B}\ }\textbf {\bibinfo {volume}
  {98}},\ \bibinfo {pages} {165129} (\bibinfo {year} {2018})},\ \bibinfo {note}
  {(see the Supplementary materials)}\BibitemShut {NoStop}%
\bibitem [{\citenamefont {Altshuler}\ and\ \citenamefont
  {Aronov}(1985)}]{Alt_Aronov}%
  \BibitemOpen
  \bibfield  {author} {\bibinfo {author} {\bibfnamefont {B.~L.}\ \bibnamefont
  {Altshuler}}\ and\ \bibinfo {author} {\bibfnamefont {A.~A.}\ \bibnamefont
  {Aronov}},\ }\href@noop {} {\emph {\bibinfo {title} {{Electron-Electron
  Interactions in Disordered Systems}}}},\ edited by\ \bibinfo {editor}
  {\bibfnamefont {A.~L.}\ \bibnamefont {Efros}}\ and\ \bibinfo {editor}
  {\bibfnamefont {M.}~\bibnamefont {Pollak}}\ (\bibinfo  {publisher} {Elsevier
  Science Publishers, Amsterdam},\ \bibinfo {year} {1985})\BibitemShut
  {NoStop}%
\bibitem [{\citenamefont {Akkermans}\ and\ \citenamefont
  {Montambaux}(2007)}]{montaumbaux}%
  \BibitemOpen
  \bibfield  {author} {\bibinfo {author} {\bibfnamefont {E.}~\bibnamefont
  {Akkermans}}\ and\ \bibinfo {author} {\bibfnamefont {G.}~\bibnamefont
  {Montambaux}},\ }\href@noop {} {\emph {\bibinfo {title} {{Mesoscopic physics
  of electrons and photons}}}}\ (\bibinfo  {publisher} {Cambridge University
  Press, Cambridge, UK},\ \bibinfo {year} {2007})\BibitemShut {NoStop}%
\bibitem [{\citenamefont {McCann}\ \emph {et~al.}(2006)\citenamefont {McCann},
  \citenamefont {Kechedzhi}, \citenamefont {Fal'ko}, \citenamefont {Suzuura},
  \citenamefont {Ando},\ and\ \citenamefont
  {Altshuler}}]{PhysRevLett.97.146805}%
  \BibitemOpen
  \bibfield  {author} {\bibinfo {author} {\bibfnamefont {E.}~\bibnamefont
  {McCann}}, \bibinfo {author} {\bibfnamefont {K.}~\bibnamefont {Kechedzhi}},
  \bibinfo {author} {\bibfnamefont {V.~I.}\ \bibnamefont {Fal'ko}}, \bibinfo
  {author} {\bibfnamefont {H.}~\bibnamefont {Suzuura}}, \bibinfo {author}
  {\bibfnamefont {T.}~\bibnamefont {Ando}}, \ and\ \bibinfo {author}
  {\bibfnamefont {B.~L.}\ \bibnamefont {Altshuler}},\ }\bibfield  {title}
  {\enquote {\bibinfo {title} {{Weak-Localization Magnetoresistance and Valley
  Symmetry in Graphene}},}\ }\href {\doibase 10.1103/PhysRevLett.97.146805}
  {\bibfield  {journal} {\bibinfo  {journal} {Phys. Rev. Lett.}\ }\textbf
  {\bibinfo {volume} {97}},\ \bibinfo {pages} {146805} (\bibinfo {year}
  {2006})}\BibitemShut {NoStop}%
\bibitem [{\citenamefont {Morozov}\ \emph {et~al.}(2006)\citenamefont
  {Morozov}, \citenamefont {Novoselov}, \citenamefont {Katsnelson},
  \citenamefont {Schedin}, \citenamefont {Ponomarenko}, \citenamefont {Jiang},\
  and\ \citenamefont {Geim}}]{PhysRevLett.97.016801}%
  \BibitemOpen
  \bibfield  {author} {\bibinfo {author} {\bibfnamefont {S.~V.}\ \bibnamefont
  {Morozov}}, \bibinfo {author} {\bibfnamefont {K.~S.}\ \bibnamefont
  {Novoselov}}, \bibinfo {author} {\bibfnamefont {M.~I.}\ \bibnamefont
  {Katsnelson}}, \bibinfo {author} {\bibfnamefont {F.}~\bibnamefont {Schedin}},
  \bibinfo {author} {\bibfnamefont {L.~A.}\ \bibnamefont {Ponomarenko}},
  \bibinfo {author} {\bibfnamefont {D.}~\bibnamefont {Jiang}}, \ and\ \bibinfo
  {author} {\bibfnamefont {A.~K.}\ \bibnamefont {Geim}},\ }\bibfield  {title}
  {\enquote {\bibinfo {title} {{Strong Suppression of Weak Localization in
  Graphene}},}\ }\href {\doibase 10.1103/PhysRevLett.97.016801} {\bibfield
  {journal} {\bibinfo  {journal} {Phys. Rev. Lett.}\ }\textbf {\bibinfo
  {volume} {97}},\ \bibinfo {pages} {016801} (\bibinfo {year}
  {2006})}\BibitemShut {NoStop}%
\bibitem [{\citenamefont {Wang}\ \emph {et~al.}(2013)\citenamefont {Wang},
  \citenamefont {Jiang}, \citenamefont {Yan}, \citenamefont {Sun},
  \citenamefont {Li}, \citenamefont {Shi},\ and\ \citenamefont
  {Chen}}]{Wang_2013}%
  \BibitemOpen
  \bibfield  {author} {\bibinfo {author} {\bibfnamefont {X.}~\bibnamefont
  {Wang}}, \bibinfo {author} {\bibfnamefont {H.~T.}\ \bibnamefont {Jiang}},
  \bibinfo {author} {\bibfnamefont {C.}~\bibnamefont {Yan}}, \bibinfo {author}
  {\bibfnamefont {Y.}~\bibnamefont {Sun}}, \bibinfo {author} {\bibfnamefont
  {Y.~H.}\ \bibnamefont {Li}}, \bibinfo {author} {\bibfnamefont {Y.~L.}\
  \bibnamefont {Shi}}, \ and\ \bibinfo {author} {\bibfnamefont
  {H.}~\bibnamefont {Chen}},\ }\bibfield  {title} {\enquote {\bibinfo {title}
  {{Anomalous transmission of disordered photonic graphenes at the Dirac
  point}},}\ }\href {\doibase 10.1209/0295-5075/103/17003} {\bibfield
  {journal} {\bibinfo  {journal} {Europhys. Lett.}\ }\textbf {\bibinfo {volume}
  {103}},\ \bibinfo {pages} {17003} (\bibinfo {year} {2013})}\BibitemShut
  {NoStop}%
\bibitem [{\citenamefont {Deng}\ \emph {et~al.}(2014)\citenamefont {Deng},
  \citenamefont {Sun}, \citenamefont {Wang}, \citenamefont {Xue}, \citenamefont
  {Li}, \citenamefont {Jiang}, \citenamefont {Shi}, \citenamefont {Chang},\
  and\ \citenamefont {Chen}}]{valley_dependent_beams}%
  \BibitemOpen
  \bibfield  {author} {\bibinfo {author} {\bibfnamefont {F.}~\bibnamefont
  {Deng}}, \bibinfo {author} {\bibfnamefont {Y.}~\bibnamefont {Sun}}, \bibinfo
  {author} {\bibfnamefont {X.}~\bibnamefont {Wang}}, \bibinfo {author}
  {\bibfnamefont {R.}~\bibnamefont {Xue}}, \bibinfo {author} {\bibfnamefont
  {Y.}~\bibnamefont {Li}}, \bibinfo {author} {\bibfnamefont {H.}~\bibnamefont
  {Jiang}}, \bibinfo {author} {\bibfnamefont {Y.}~\bibnamefont {Shi}}, \bibinfo
  {author} {\bibfnamefont {K.}~\bibnamefont {Chang}}, \ and\ \bibinfo {author}
  {\bibfnamefont {H.}~\bibnamefont {Chen}},\ }\bibfield  {title} {\enquote
  {\bibinfo {title} {{Observation of valley-dependent beams in photonic
  graphene}},}\ }\href {\doibase 10.1364/OE.22.023605} {\bibfield  {journal}
  {\bibinfo  {journal} {Optics Express}\ }\textbf {\bibinfo {volume} {22}},\
  \bibinfo {pages} {23605} (\bibinfo {year} {2014})}\BibitemShut {NoStop}%
\bibitem [{\citenamefont {Kechedzhi}\ \emph {et~al.}(2008)\citenamefont
  {Kechedzhi}, \citenamefont {Kashuba},\ and\ \citenamefont
  {Fal'ko}}]{PhysRevB.77.193403}%
  \BibitemOpen
  \bibfield  {author} {\bibinfo {author} {\bibfnamefont {K.}~\bibnamefont
  {Kechedzhi}}, \bibinfo {author} {\bibfnamefont {O.}~\bibnamefont {Kashuba}},
  \ and\ \bibinfo {author} {\bibfnamefont {V.~I.}\ \bibnamefont {Fal'ko}},\
  }\bibfield  {title} {\enquote {\bibinfo {title} {{Quantum kinetic equation
  and universal conductance fluctuations in graphene}},}\ }\href {\doibase
  10.1103/PhysRevB.77.193403} {\bibfield  {journal} {\bibinfo  {journal} {Phys.
  Rev. B}\ }\textbf {\bibinfo {volume} {77}},\ \bibinfo {pages} {193403}
  (\bibinfo {year} {2008})}\BibitemShut {NoStop}%
\end{thebibliography}%
\end{document}